\documentclass[aps,showpacs,preprintnumbers,nofootinbib,amsmath,amssymb]{revtex4-1}
\usepackage{amssymb}
\usepackage{}
\usepackage{slashed,color,amsmath,amssymb,mathrsfs}
\usepackage{graphicx}
\usepackage{ulem}
\usepackage{hyperref}
\usepackage{longtable}
\usepackage{array}
\usepackage{bm}

\allowdisplaybreaks

\begin{document}
\preprint{}
\title{Coupled-Channel-Induced $S-D$ mixing of Charmonia and Testing Possible Assignments for $Y(4260)$ and $Y(4360)$}
\author{Hui-Feng Fu}
\email{huifengfu@jlu.edu.cn}
\affiliation{Center for Theoretical Physics, College of Physics, Jilin University,
Changchun 130012, P. R. China}
\author{Libo Jiang}
\email{jiangl@fnal.gov}
\affiliation{Department of Physics and Astronomy, University of Pittsburgh, Pittsburgh, PA 15260, USA}
\begin{abstract}
{
The mass spectrum and the two-body open-charm decays of the $J^{PC}=1^{--}$ charmonium states are studied with the coupled-channel effects taken into account. The coupled-channel-induced mixing effects among the excited vector charmonia are studied. Based on our calculations of the masses and the decay widths, we find that the tension between the observed properties of $Y(4260)/Y(4360)$ and their conventional charmonia interpretations could be softened.
}
\end{abstract}
 \maketitle
\section{Introduction}

In 2005, { the BaBar Collaboration} observed a broad structure, $Y(4260)$, near 4.26 GeV in the initial state radiation $e^+e^- \rightarrow \gamma_{\mathrm{ISR}}\pi^+\pi^- J/\psi$ process~\cite{4260}. {Later, it was confirmed by the CLEO~\cite{CLEO42} and Belle~\cite{Belle42} Collaborations. This state is considered as a non-$q\bar{q}$ state by many authors, such as the hybrid state~\cite{Zhu1,Kou}, the tetraquark state~\cite{Mai,Ebert,ZGWang1,Ali}, the molecular state~\cite{Yuan,Ding2,Close,Cleven}, the hadrocharmonium~\cite{DubyS} and so on,} because its mass and decay properties seem to be in conflict with expectations. Conventionally, $\psi(4040)$, $\psi(4160)$ and $\psi(4415)$ are usually assigned as the $\psi(3S)$, $\psi(2D)$ and $\psi(4S)$ states, respectively, so a resonance with the mass near 4.26 GeV cannot fit in. In addition, the already confirmed excited charmonia usually decay into open-charm states dominantly once they exceed the open-charm threshold, but $Y(4260)$ has never been found in { two-body open-charm decay channels~\cite{PDG}\footnote{Recently, the BESIII Collaboration reported the observation of a resonance at $4228.6\pm4.1\pm6.3$ MeV in the $e^+e^- \rightarrow \pi^+D^0D^{*-}$ process~\cite{BESopen}. This is the first observation of the $Y$ state in the open-charm channels.}}.
Another resonance, $Y(4360)$, is in a quite similar situation as $Y(4260)$. { It was first observed by the BaBar Collaboration in 2007~\cite{4360}, and was confirmed by the Belle Collaboration soon~\cite{Belle43}. It is also considered as a good non-$q\bar{q}$ candidate by many researchers~\cite{Kal,CloseF2,MaianiL,ZGWang2,QiaoC2,Qiao,Vol}. (We refer the readers to Refs.~\cite{CCLZ,Esposito,Lebed,Ali2,Guo,LiuY} for a comprehensive review on $Y(4260)$ and $Y(4360)$.)}

{ Although the masses of $Y(4260)$ and $Y(4360)$ are usually considered to be incompatible with the conventional charmonia picture,  several studies predicted relatively compact mass spectra, with the masses of $\psi(4S)$ and/or $\psi(3D)$ in the $~4.2-4.4$ GeV region. In Refs.~\cite{Shah,Zhang}, $Y(4260)$ is interpreted as a conventional charmonium. In Refs.~\cite{Ding,Anwar,Seg}, $Y(4360)$ is considered to be a charmonium. And in Ref.~\cite{Li}, $Y(4260)$ and $Y(4360)$ are assigned as $\psi(4S)$ and $\psi(3D)$ respectively. For the decay properties, since the excited vector charmonia are usually mixtures of the $^3S_1$ and $^3D_1$ states, the author of Ref.~\cite{Lla} suggested that the non-observation of $Y(4260)$ in $e^+e^-\rightarrow$ hadrons may be due to the interference between the $S$-wave and $D$-wave contributions. (We shall call the mixing among the $S$-wave and $D$-wave vector charmona the $S-D$ mixing for simplicity.) In addition, highly-excited charmonia may have unexpected decay properties due to their node structures. Thus, the common arguments on rejecting conventional interpretations of $Y(4260)/Y(4360)$ may not stand up firmly. So, even though many studies favor exotic hadronic interpretations for these states, since no commonly accepted conclusions have occurred, we think it is still very interesting to investigate the mass spectrum and open-charm decays of vector charmonia incorporating the $S-D$ mixing effect, in order to see whether the tension between the observed properties of $Y(4260)/Y(4360)$ and their conventional charmonia interpretations can be softened, and to test particular assignments for $Y(4260)/Y(4360)$ in the conventional charmonium picture.

Such an investigation should incorporate the coupled-channel effects~\cite{Cor1,Cor2}, because they have significant impacts on the excited charmonia, especially, the coupled-channel effects induce the $S-D$ mixing of vector charmona. In coupled-channel models, the quarkonium couples to the continuum multi-particle states, and the physical states are mixtures among all these states. Many studies on coupled-channel effects in the vector charmonium sector focused on the mass shifts and ignored the $S-D$ mixing, such as Refs.~\cite{YS,Chao}.}  The coupled-channel effects with all possible mixtures were studied in Refs.~\cite{Cor1,Cor2,Ono1,Ono2,Eich},either using the Cornell model or using {the $^3P_0$ model.} For vector charmonia, it is expected that the mixing between the $(n+1)S$ and $nD$ states is most significant compared to the mixing among other states, which was justified in Refs.~\cite{Ono1}. {So, in this work,} only the mixing between the $(n+1)S$ and $nD$ states are considered. The decay channels we studied are $D^{(*)}\bar{D}^{(*)}$, $D_s^{(*)}\bar{D}_s^{(*)}$. { We assume $Y(4260)$ and $Y(4360)$ to be the lower and higher states of the mixture of $\psi(4S)$ and $\psi(3D)$ respectively, and compare the results with experimental data to test such an assumption. In addition, the coupled-channel-induced $S-D$ mixing of vector charmonia on its own has physical significants, because it could be tested with experiments through di-leptonic decays.}

{This paper is organized as follows.} The theoretical tools are introduced in Section \uppercase\expandafter{\romannumeral2}. We use the instantaneous Bethe-Salpeter equation with the Cornell potential to calculate the wave functions of relevant mesons. Then the $^3P_0$ model is used to evaluate decay amplitudes. The coupled-channel dynamics is reviewed in this section. {In Section \uppercase\expandafter{\romannumeral3}, the results and discussions are presented.} The last section is devoted to conclusions.

\section{Theoretical settings}
\subsection{Coupled-Channel Dynamics}
In naive quark models, mesons are bound states of a quark and an anti-quark bounded by a QCD-inspired potential. The masses and wave functions can be obtained by solving the eigenvalue problem formally expressed as:
\begin{equation}\label{H0}
H_0|\psi_n\rangle=M_{0,n} |\psi_n\rangle,
\end{equation}
where the $M_0$ and $|\psi_n\rangle$ are usually referred as the bare mass and the bare state. The Hamiltonian $H_0$ only includes the interaction described by the potential binding the quark-anti-quark pair. In coupled-channel formulism, the Hilbert space under consideration is enlarged to include the continuum states which the bare states can decay into. We are focusing on vector charmona decaying into two-body open-charm channels, and in this situation, the Hamiltonian can be formally written as~\cite{Ono2}:
\begin{equation}\label{Htotal}
H=\left\{
\begin{array}{cc}
H_0&H_{\mathrm{QPC}}^\dag\\
H_{\mathrm{QPC}}&H_{BC}
\end{array}
\right\},
\end{equation}
where $H_{BC}$ is the free Hamiltonian (by "free", we mean that the interactions between the two mesons are neglected) for the continuum states with two particles $B$ and $C$:
\begin{equation}\label{HBC}
H_{BC}|B,C;\bm{P}_B,\bm{P}_C\rangle=E_{BC}|B,C;\bm{P}_B,\bm{P}_C\rangle,
\end{equation}
where
\begin{equation}
E_{BC}=\sqrt{M_B^2+\bm{P}_B^2}+\sqrt{M_C^2+\bm{P}_C^2}.
\end{equation}
The quark-pair creation Hamiltonian $H_{\mathrm{QPC}}$ induces the decays. With the presence of the non-diagonal elements in the Hamiltonian, the physical states become mixtures of all bare states:
\begin{equation}
|\psi'_n\rangle= \sum_{i} a_{ni} |\psi_i\rangle+ \sum_{BC} \int dk~c_{n,BC} |B,C;\bm{P}_B,\bm{P}_C\rangle,
\end{equation}
where $\int dk$ denotes an integration over all three-momenta of $B$ and $C$.
The problem now turns into the eigenvalue problem~\cite{Cor1,Cor2}:
\begin{equation}\label{Mass}
\det |(M-M_{0,n})\delta_{mn}-\Pi_{mn}(M)|=0,
\end{equation}
where
\begin{equation}
\Pi_{mn}(M)=\sum_{BC}\int dk \frac{\langle m|H_{\mathrm{QPC}}^\dag|B,C;\bm{P}_B,\bm{P}_C\rangle\langle B,C;\bm{P}_B,\bm{P}_C|H_{\mathrm{QPC}}|n\rangle}{M-E_{BC}+i\epsilon}.
\end{equation}
Above the open-charm threshold, $\Pi_{mn}$ develops an imaginary part, so in general, the equation allows complex value solutions. In this case, the physical mass and the width of a physical state $|\psi'_n\rangle$ are related to the corresponding eigenvalue $M_n$ as
\begin{eqnarray}
M_{\mathrm{phys.},n}&=&\mathrm{Re} M_n,\\
\Gamma_n &=& -2~\mathrm{Im} M_n.
\end{eqnarray}
For real $M_n$'s, of cause, $M_n$ is just the physical mass of the corresponding state, and these $M_n$'s should satisfy the condition $M_n<2M_D$, where $2M_D$ denotes the open-charm threshold.

In vector charmonium sector, the mass of $nD$ state is closest to the $(n+1)S$ state. So it is expected that these two states should mix each other the most, which is verified in Ref.~\cite{Ono1} and taken as granted by many authors~\cite{Bad,Zou}. In this work, we follow these researches and only take into account the mixing between the $nD$ and the $(n+1)S$ states, which means $\Pi_{nD,(n+1)S}$ and $\Pi_{(n+1)S,nD}$ are the only non-vanishing non-diagonal elements. With this simplification, Eq. (\ref{Mass}) decomposes into several sectors. For $1S$ sector, we have
\begin{equation}\label{1SMass}
(M-M_{0,1})-\Pi_{1S1S}(M)=0.
\end{equation}
For $(n+1)S-nD$ sectors, we have
\begin{equation}\label{SDMass}
\det \left|
\begin{array}{cc}
M-M_{0,S}-\Pi_{SS}(M)&-\Pi_{SD}(M)\\
-\Pi_{DS}(M)&M-M_{0,D}-\Pi_{DD}(M)
\end{array}
\right|=0,
\end{equation}
where the principle quantum numbers in front of $S$ and $D$ are omitted. Since we are focusing on the $S-D$ mixing in this work, and the continuum components in the physical states are irrelevant, we drop off these components in the physical states as Refs.~\cite{Ono1,Zou} did. So a mixed physical state is expressed as
\begin{equation}\label{SDa}
|\psi'\rangle=a_S|S\rangle+a_D|D\rangle,
\end{equation}
with coefficients $a_S$ and $a_D$ satisfying $|a_S|^2+|a_D|^2=1$.
\subsection{Instantaneous Bethe-Salpeter Equation and the $^3P_0$ Model}
Now the remaining problem is to solve the naive quark model to obtain the bare masses and the wave functions, and to calculate the open-charm decay amplitudes $\langle B,C;\bm{P}_B,\bm{P}_C|H_{\mathrm{QPC}}|n\rangle$. To this end, we make use of the instantaneous Bethe-Salpeter equation and the $^3P_0$ model. The instantaneous BS equation, also known as the Salpeter equation, is a well developed relativistic two-body bound state equation, and is very suitable to apply on the heavy quarkonium system. The instantaneous BS wave function $\varphi(q_{_{P_\perp}}^{\mu})$ may be decomposed into positive and negative energy parts: $\varphi=\varphi^{++}+\varphi^{--}$, where $q_{_{P_\perp}}$ is the perpendicular part of relative momentum $q$. For any momentum $l^\mu$, we have $l_{_{P_\perp}}=l-\frac{l_{_P}}{\sqrt{P^2}}P$ and $l_{_P}\equiv\frac{l\cdot P}{\sqrt{P^2}}$, where $P^\mu$ is the meson's momentum. The Salpeter equation then takes the form as coupled equations for $\varphi^{++}$ and $\varphi^{--}$~\cite{Salpeter}:
\begin{eqnarray}\label{Salpeter}
&&(M_0-\omega_{1_P}-\omega_{2_P})\varphi^{++}(q_{_{P_\perp}})=\Lambda_{1}^{+}(p_{1_{P_\perp}})\eta(q_{_{P_\perp}})
\Lambda_{2}^{+}(p_{2_{P_\perp}}),\notag\\
&&(M_0+\omega_{1_P}+\omega_{2_P})\varphi^{--}(q_{_{P_\perp}})=-\Lambda_{1}^{-}(p_{1_{P_\perp}})\eta(q_{_{P_\perp}})
\Lambda_{2}^{-}(p_{2_{P_\perp}}),
\end{eqnarray}
where
\begin{eqnarray}
\Lambda_{j}^{\pm}(p_{j_{P_\perp}})&\equiv&
\frac{1}{2\omega_{j_P}}[\frac{\slashed
P}{\sqrt{P^2}}\omega_{j_P}\pm(\slashed
p_{j_{P_\perp}}+(-1)^{j+1}m_{j})],\\
\eta(q_{_{P_\perp}}^{\mu})&\equiv& \int
\frac{\mathrm{d}k_{_{P_\perp}}^{3}}{(2\pi)^{3}}V(k_{_{P_\perp}}^{\mu}-q_{_{P_\perp}}^{\mu})\varphi(k_{_{P_\perp}}^{\mu}),\\
\omega_{j_P}&\equiv &\sqrt{m_{j}^{2}-p_{j_{P_\perp}}^{2}},
\end{eqnarray}
with $j=1$ for quark and $j=2$ for anti-quark. $p_1$ is the quark momentum and $p_2$ is the anti-quark momentum. The instantaneous interaction kernel $V(k_{_{P_\perp}}^{\mu}-q_{_{P_\perp}}^{\mu})$ now becomes the QCD-inspired potential between quark and anti-quark. In our model, we use a modified Cornell potential, which in the center of mass frame reads~\cite{Chang,Wang}:
\begin{eqnarray}
&\displaystyle V(\bm{q})=V_s(\bm{q})+V_v(\bm{q})\gamma^0\otimes\gamma_0,\notag\\
&\displaystyle
V_s(\bm{q})=-(\frac{\lambda}{\alpha}+V_0)\delta^3(\bm{q})
+\frac{\lambda}{\pi^2}\frac{1}{(\bm{q}^2+\alpha^2)^2},\notag\\
&\displaystyle
V_v(\bm{q})=-\frac{2}{3\pi^2}\frac{\alpha_s(\bm{q})}{(\bm{q}^2+\alpha^2)},\notag
\end{eqnarray}
where $\bm{q}$ is the three-momentum of $q_{_{P_\perp}}^{\mu}$, i.e., $q_{_{P_\perp}}^{\mu}=(0,\bm{q})$ in the meson's rest frame.
$\alpha_s(\bm{q})=\frac{12\pi}{33-2N_f}\frac{1}{\mathrm{log}(a+\bm{q}^2/\Lambda_{\mathrm{QCD}}^2)}$ is the QCD running coupling constant;
the constants $\lambda,\ \alpha,\ a,\ V_0$ and
$\Lambda_{\mathrm{QCD}}$ are the parameters characterizing the
potential.

The method of solving the full Salpeter equation is given in Ref.~\cite{Chang,Wang}. After solving the Salpeter equation, we obtain the bare mass spectrum and the wave functions. Then we can use them to calculate the open-charm decay amplitudes. The $A\rightarrow BC$ open-charm decay is a typical OZI-allowed strong decay process, which in essence is a non-perturbative QCD problem. Due to our pure knowledge of QCD in its non-perturbative region, such processes are usually evaluated using models. Among others, the $^3P_0$ model is a widely accepted one~\cite{Micu,Le}. This model assumes the decay takes place via creating a quark anti-quark pair from the vacuum carrying the quantum number $^3P_0$, so it is also known as the quark pair creation model. The $^3P_0$ model is a non-relativistic model, and majority of works using this model stick on its non-relativistic form. On the other hand, the Salpeter wave function is a relativistic wave function containing the Dirac spinor of quark/anti-quark. So incorporating the Salpeter wave function into $^3P_0$ model requires the relativistic extension of the original $^3P_0$ model, which has been done in Ref.~\cite{Fu}. In this work we employ the formula derived in that reference and write the amplitude for the $A\rightarrow BC$ open-charm decay as
\begin{equation}\label{eq9}
\mathcal{M}=g\int
\frac{d^3q^A_{_{P^A_\perp}}}{(2\pi)^3}\mathrm{Tr}\{\frac{\slashed
P_A}{M_A}\varphi^{++}_A(q^A_{_{P^A_\perp}})\frac{\slashed
P_A}{M_A}\bar{\varphi}^{++}_C(q^C_{_{P^A_\perp}})\bar{\varphi}^{++}_B(q^B_{_{P^A_\perp}})\},
\end{equation}
where quantities referred to $A$, $B$ and $C$ are labeled with the subscript/superscript $A$, $B$ and $C$ respectively. $g=2m_q \gamma$ with $m_q$ being the the mass of the created quark $q$ or anti-quark $\bar{q}$ and $\gamma$ being a universal constance characterizing the strength of the decay. Negative-energy contributions have been neglected due to their smallness comparing to positive-energy contributions.

For $J^{P}=1^{-}$ meson, the wave function $\varphi^{++}$ can be decomposed into two parts: the $S$-wave part and the $D$-wave part, i.e.,
\begin{equation}\label{eq19}
\varphi^{++}_{1^-}(q_{_{P_\perp}})=\sqrt{\frac{m_1m_2}{\omega_{1_P}\omega_{2_P}}}\sum_{s_1,s_2,m,\tilde{M}}u_{s_1}(\bm{p}_1)\bar{v}_{s_2}(\bm{p}_2)\chi^{1,\tilde{M}}_{s_1,s_2}
(\sqrt{2M_04\pi}R_{n0}Y_{00}\delta_{m,0}\delta_{\tilde{M},\lambda}+\sqrt{2M_04\pi}R_{n2}Y_{2m}S_{m,\tilde{M};1,\lambda}^{2,1}),
\end{equation}
where $R_{nl}$ is the radial wave function with principle number $n$ and orbital angular momentum $l$. $Y_{lm}$ is the spherical harmonic function. $m$ is the magnetic quantum number of orbital angular momentum. $S,\tilde{M}$ represent the total spin of the quark-anti-quark pair. The $\lambda$ appearing in Eq. (\ref{eq19}) is the magnetic quantum number of the meson. $s_{1(2)}$ is the
spin of the quark (anti-quark) in the meson.
$S_{m,\tilde{M};J,M_{J}}^{l,S}=\langle \tilde{M},m|J,
M_{J}\rangle$ and $\chi_{s_1,s_{2}}^{S,\tilde{M}}=\langle
s_1,s_{2}|S, \tilde{M}\rangle$ are Clebsch-Gordan (C-G)
coefficients. $u_{s}(\bm{p})$ ($\bar{v}_{s}(\bm{p})$) is the Dirac spinor of the quark (anti-quark) with spin $s$ and momentum $\bm{p}$. This expression clearly shows the $L-S$ coupling inside the meson, and it is the same as the corresponding non-relativistic wave function except that: a) the non-relativistic spinor is replaced with the Dirac spinor; b) the radial wave function is the solution of the Salpeter equation rather than the non-relativistic Schr$\mathrm{\ddot{o}}$dinger equation.

Inserting the $S$-wave part or the $D$-wave part of the wave function into Eq. (\ref{eq9}), one arrives at the familiar form of the decay amplitude of the $^3P_0$ model:
\begin{eqnarray}\label{eq23}
\mathcal{M}&=&g\int
\frac{d^3q_{_{P_\perp}}}{(2\pi)^3}\{\sum_{\tilde{M},m}\sum_{\tilde{M}_{B},m_B}\sum_{\tilde{M}_{C},m_C}\psi_{nlm}(q_{_{P_\perp}})(\psi_{n_Cl_Cm_C}(q^C_{_{P_\perp}}))^*
(\psi_{n_Bl_Bm_B}(q^B_{_{P_\perp}}))^*\notag\\
&&\times
S_{m,\tilde{M};J,M_{J}}^{l,S}S_{m_C,\tilde{M}_C;J_C,M_{J_C}}^{l_C,S_C}S_{m_B,\tilde{M}_B;J_B,M_{J_B}}^{l_B,S_B}
\sum_{s_a,s_{\bar{a}},s_q,s_{\bar{q}}}\chi^{S,\tilde{M}}_{s_a,s_{\bar{a}}}\chi^{S_C,\tilde{M}_C}_{s_q,s_{\bar{a}}}\chi^{S_B,\tilde{M}_B}_{s_a,s_{\bar{q}}}\notag\\
&&\times
\frac{m_q}{\omega_{q_P}}\bar{u}_{s_q}(\bm{p}_q)v_{s_{\bar{q}}}(\bm{p}_{\bar{q}}),
\end{eqnarray}
where $\psi_{nlm}=\sqrt{2M_04\pi}R_{nl}Y_{lm}$ and $\omega_{q_P}=\sqrt{\bm{p}_q^2+m_q^2}$. The factor $\frac{m_q}{\omega_{q_P}}$ is usually taken to be 1 in a non-relativistic $^3P_0$ model calculation, however in this work, we keep this factor as it is and serve it as a relativistic correction in the decay amplitudes. The decay width is given by
\begin{equation}
\Gamma=\frac{|\bm{P}_B|}{8M_A^2\pi}\frac{1}{3}\sum_{\mathrm{pol}}|\mathcal{M}|^2
=\frac{2\pi|\bm{P}_B|E_BE_C}{M_A}\frac{1}{3}\sum_{\mathrm{pol}}|f_{A\rightarrow BC}|^2,
\end{equation}
where $f_{A\rightarrow BC}$ is introduced to make connection with the convention which is widely used in literature (such as Ref.~\cite{Swanson}). $\sum_{\mathrm{pol}}$ means summation over polarizations. $f_{A\rightarrow BC}$ is related to the matrix element as
\begin{equation}
\langle B,C;\bm{P}_B,\bm{P}_C|H_{\mathrm{QPC}}|A\rangle= \delta^3(\bm{P}_A-\bm{P}_B-\bm{P}_C)f_{A\rightarrow BC}.
\end{equation}

Given these decay amplitudes, whenever $\Pi_{mn}$ develops imaginary part, $\Pi_{mn}$ can be calculated as~\cite{Ono2}
\begin{eqnarray}
\mathrm{Im} \Pi_{mn} (E)&=&-\sum_{BC}\left\{\frac{1}{3}\sum_{\mathrm{pol}} f_{m\rightarrow BC}^* f_{n\rightarrow BC}\times\frac{\pi|\bm{P}_B|E_BE_C}{E} \theta(E-E_{BC}^{th})\right\},\\
\mathrm{Re} \Pi_{mn} (E)&=&-\frac{1}{\pi}\int dE' \frac{\mathrm{Im} \Pi_{mn} (E')}{E-E'},
\end{eqnarray}
where $E_{BC}^{th}$ represents the threshold energy of $BC$ channel. $\theta(x)$ is the step function. The expressions of $\mathrm{Im} \Pi_{mn}$ for each decay channels are given in Appendix A.

\section{Results and Discussions}
Previous section gives the formula for calculating the charmonium spectrum and the coupled-channel effects. To obtain the bare masses and the bare states of vector charmonia, we use the following parameters:
\begin{equation}
\begin{array}{ccc}
m_c=1.71~\mathrm{GeV},~~~ & a=\mathrm{e}=2.7183, &~~~\alpha=0.12~ \mathrm{GeV},\notag\\
\lambda=0.202~ \mathrm{GeV}^2,~~~& \Lambda_{\mathrm{QCD}}=0.40~ \mathrm{GeV}, & ~~~V_0=-0.204~ \mathrm{GeV}.\notag
\end{array}
\end{equation}
While all the other parameters lie in reasonable ranges, the parameter $\Lambda_{\mathrm{QCD}}$ is a bit larger than its typical value. However, first of all, $\Lambda_{\mathrm{QCD}}$ is a renormalization-scheme-dependent parameter, and its value varies from scheme to scheme. Secondly, in the popular $\overline{MS}$ scheme, at two-loop order, $\Lambda_{\overline{MS}}^{(4)}=325$~MeV corresponds to $\alpha_s(M_z)=0.118$, where the superscript of $\Lambda_{\overline{MS}}^{(4)}$ represents $n_f=4$, while, $\Lambda_{\overline{MS}}^{(4)}=413$~MeV corresponds to $\alpha_s(M_z)=0.123$~\cite{Buras}. So, the value we take for $\Lambda_{\mathrm{QCD}}$ is not unreasonable large. Moreover, we have checked that varying the parameter $\Lambda_{\mathrm{QCD}}$ has impacts on the mass spectrum, but much less impacts on the $S-D$ mixing angles and the decay widths.

\subsection{The mass spectrum}
The results of bare masses are listed in Table \ref{tab1}. With the obtained wave functions, the coupled-channel effects are calculated. The decay strength parameter of $^3P_0$ model is fitted to be $\gamma=0.43$, and we also set the strength parameter of creating $s\bar{s}$ to be $\gamma_s=\frac{\gamma}{\sqrt{3}}$ as usual~\cite{Le2,Liu}. For consistency, the wave functions of $D^{(*)}$ and $D^{(*)}_s$ are also calculated with the Salpeter equation. With constituent quark mass $m_u=m_d=0.305$ GeV and $m_s=0.500$ GeV, we obtain $M_{D}=1.865$ GeV, $M_{D^*}=2.008$ GeV, $M_{D_s}=1.968$ GeV and $M_{D_s^*}=2.112$ GeV. Now the physical states are mixtures of the bare $\psi((n+1)S)$ and $\psi(nD)$ states. For $\psi(1S)$, we assume that it does not mix with other bare states. We denote the physical state as $\psi'(nS)/\psi'(nD)$, where $nS/nD$ indicates the dominant component in this physical state. The physical masses are shifted by the coupled-channel effects and can be compared with the experimental data. The mass shifts are represented by $\Delta M$. We also list some results from literature for comparison. These information are gathered in Table \ref{tab1}. One can see that from the $\psi'(1S)$ through $\psi'(3S)$, the masses are comparable with the experimental data, which justifies our model calculations. ($\psi'(2D)$ is one exception, which is also a problem of another potential model calculation~\cite{Li}.)

We compare the masses of the two physical states $\psi'(4S)$ and $\psi'(3D)$ with the masses of $Y(4260)$ and $Y(4360)$ respectively. The mass of $Y(4260)$ was measured by BaBar~\cite{4260} and Belle~\cite{Belle} both close to 4260 MeV, but the most recent measurement performed by BES\uppercase\expandafter{\romannumeral3} gives a mass of $4222.0\pm 3.1\pm 1.4$ MeV~\cite{BES3}. So the world averaged mass of $Y(4260)$ now becomes $4230\pm8$ MeV~\cite{PDG}. Our calculated mass of $\psi'(4S)$ is larger than this value by $\sim 50$ MeV. On the other hand, the mass of $\psi'(3D)$ is lower than the world averaged mass of $Y(4360)$ by $\sim 50$ MeV. We indicate that although the Cornell potential with coupled-channel effects is a good modeling for QCD dynamics, there are still some uncertainties in the mass spectrum which could be of $\sim 10$ MeV order or even larger. { We find that the masses of $\psi'(4S)$ and $\psi'(3D)$ are close to the masses of $Y(4260)$ and $Y(4360)$ respectively, yet deviations still exist. If the mixing between the $4S$ and $3D$ charmonium states is larger, the masses of $\psi'(4S)$ and $\psi'(3D)$ would be closer to the masses of $Y(4260)$ and $Y(4360)$ respectively. This possibility will be discussed in subsection C.} The bare mass of $\psi(5S)$ state in our model is 4468 MeV. Considering that the mass shifts are in general tens MeV, we find that $\psi(4415)$ should be assigned as the physical state dominated by $5S$ in our model.

\begin{table} \caption{Our theoretical results of bare masses, physical masses and mass shifts of vector charmonia. Experimental data and some theoretical predictions from literature are given for comparison. All quantities are given in the unit of MeV.}
\begin{center}\begin{tabular}{|c|c|c|c|c|c|c|}
\hline
&$M_0$ of ours&$M_{\mathrm{phys.}}$ of ours&$M$ of Ref.~\cite{Li}&$M_{\mathrm{ex.}}$~\cite{PDG}&$\Delta M$ of ours&$\Delta M$ of Ref.~\cite{Chao}\\
\hline
$\psi(1S)(\psi'(1S))$&3170&3103&3097&$3096.900\pm 0.006$&-67&-159\\
\hline
$\psi(2S)(\psi'(2S))$&3764&3666&3673&$3686.097\pm0.025$&-98&-227\\
\hline
$\psi(1D)(\psi'(1D))$&3870&3768&3787&$3773.13\pm0.35$&-102&-231\\
\hline
$\psi(3S)(\psi'(3S))$&4092&4018&4022&$4039\pm1$&-74&\\
\hline
$\psi(2D)(\psi'(2D))$&4152&4089&4089&$4191\pm5$&-63&\\
\hline
$\psi(4S)(\psi'(4S))$&4311&4285&4273&$4230\pm8$~($Y(4260)$)&-26&\\
\hline
$\psi(3D)(\psi'(3D))$&4350&4319&4317&$4368\pm13$~($Y(4360)$)&-31&\\
\hline
\end{tabular}\end{center}\label{tab1}
\end{table}

\subsection{The $S-D$ mixing angles}
In Table \ref{tab2}, we give the coefficients $a_S$ and $a_D$ as in Eq. (\ref{SDa}) for different states. We also extracted the mixing angles defined as in
\begin{equation}\label{SDmixing1}
|\psi'(S)\rangle=\cos \theta |S\rangle + \sin\theta|D\rangle,
\end{equation}
\begin{equation}\label{SDmixing2}
|\psi'(D)\rangle=-\sin \theta |S\rangle + \cos\theta|D\rangle.
\end{equation}
In extracting the mixing angles, we neglect the phase of the complex number $a_{S/D}$ and set the sign equal to the sign of its real part. The magnitude of mixing angles for $2S-1D$ states in our model are smaller than the result from Ref. \cite{Cor2} where they obtained a mixing angle of $-10^{\circ}$. The difference may due to different model settings in these two works.
The magnitude of mixing angles for $3S-2D$ states are larger than those for $2S-1D$ states as expected. For $4S-3D$ states, the mixing angles are around $-6^{\circ}\sim -5^{\circ}$.
\begin{table*} \caption{The coefficients $a_S$ and $a_D$ characterizing the $S-D$ mixing for different states and the extracted mixing angles.}
\begin{center}\begin{tabular}{|c|c|c|c|}
\hline
&$\psi(2S)$&$\psi(1D)$&$\theta_{2S-1D}$\\
\hline
$\psi'(2S)$&0.9998&-0.0194&$-1.1^{\circ}$\\
\hline
$\psi'(1D)$&$0.0754+0.0625~\mathrm{i}$&0.9952&$-5.6^{\circ}$\\
\hline
\hline
&$\psi(3S)$&$\psi(2D)$&$\theta_{3S-2D}$\\
\hline
$\psi'(3S)$&0.9697&$-0.2027+0.1360~\mathrm{i}$&$-14.1^{\circ}$\\
\hline
$\psi'(2D)$&0.2035+0.0556 &0.9775&$-12.2^{\circ}$\\
\hline
\hline
&$\psi(4S)$&$\psi(3D)$&$\theta_{4S-3D}$\\
\hline
$\psi'(4S)$&0.9956&$-0.0714 -0.0610~\mathrm{i}$&$-5.4^{\circ}$\\
\hline
$\psi'(3D)$&$0.0341+0.0950~\mathrm{i}$ &0.9949&$-5.8^{\circ}$\\
\hline
\end{tabular}\end{center}\label{tab2}
\end{table*}

\begin{figure}[hbt]
\centering
\includegraphics[width = 0.6\textwidth]{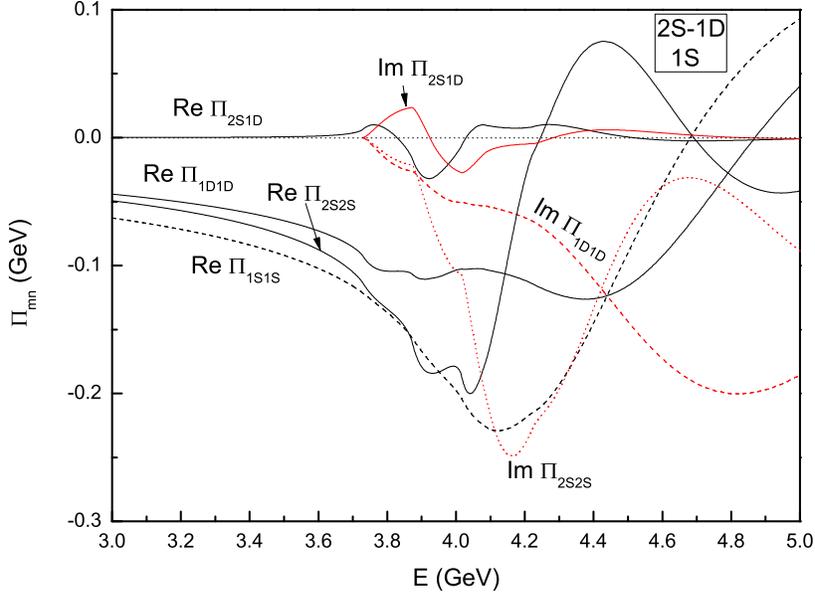}
\caption{Real and imaginary parts of $\Pi_{mn}$ in $2S-1D$ sector and Real part of $\Pi_{mn}$ in $1S$ sector.} \label{fig1}
\end{figure}

\begin{figure}[hbt]
\centering
\includegraphics[width = 0.6\textwidth]{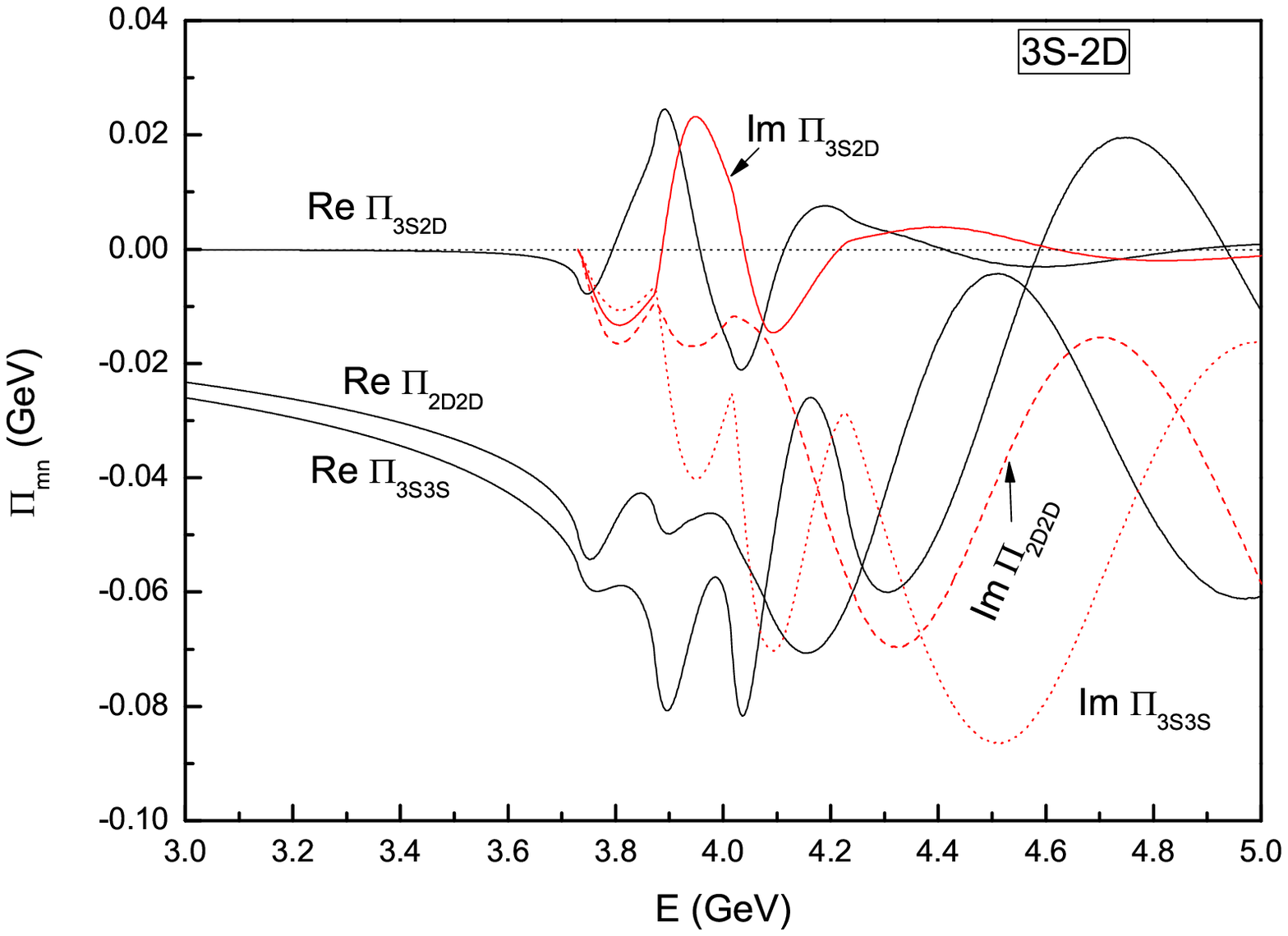}
\caption{Real and imaginary parts of $\Pi_{mn}$ in $3S-2D$ sector.} \label{fig2}
\end{figure}

\begin{figure}[hbt]
\centering
\includegraphics[width = 0.6\textwidth]{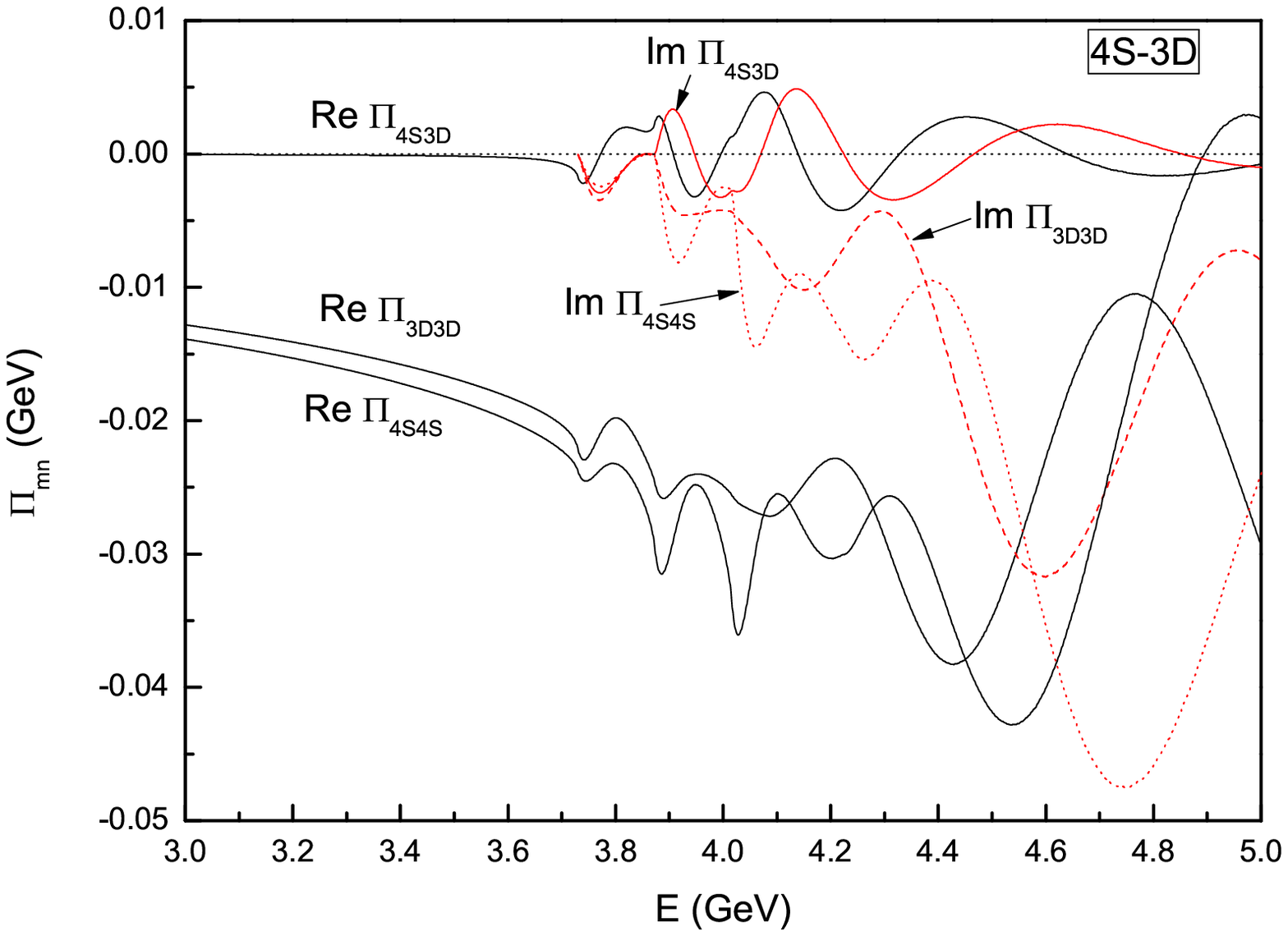}
\caption{Real and imaginary parts of $\Pi_{mn}$ in $4S-3D$ sector.} \label{fig3}
\end{figure}

To provide more details, we plot the real and imaginary parts of $\Pi_{mn}$ in Figs. \ref{fig1}-\ref{fig3}. The real parts of diagonal elements roughly reflect the mass shifts at given energy scale. The imaginary part of diagonal elements roughly reflect half-widths at given energy scale. And non-diagonal elements may reflect how much the $S-D$ mixing is at given energy scale. One can find that the non-diagonal elements are in general smaller than the diagonal ones. Actually, the diagonal elements $\mathrm{Im} \Pi_{SS}^{BC}$ and $\mathrm{Im} \Pi_{DD}^{BC}$ for each channel $BC$ always $\leq 0$, so the contributions to $\mathrm{Im}\Pi_{SS}$ and $\mathrm{Im} \Pi_{DD}$ add up. But the non-diagonal elements $\mathrm{Im} \Pi_{SD}^{BC}$ for different channels may have different signs, so $\mathrm{Im}\Pi_{SD}$ is generally smaller than $\mathrm{Im}\Pi_{SS}$ and $\mathrm{Im} \Pi_{DD}$, and oscillates around zero as $E$ varies. This in turn results in $\mathrm{Re}\Pi_{SD}<\mathrm{Re}\Pi_{SS}$ and $\mathrm{Re}\Pi_{DD}$ in general. The diagonal elements $\mathrm{Im}\Pi_{SS}$ and $\mathrm{Im}\Pi_{DD}$ also oscillate but without changing signs. The oscillation behavior of the diagonal elements of $\Pi$ is a reflection of the node structures of initial states, so this behavior becomes more frequent (against $E$) for higher excitation states.

\subsection{Decay widths}
\begin{table*} \caption{Decay widths of $D^{(*)}\bar{D}^{(*)},D_s^{(*)}\bar{D}_s^{(*)}$ channels for states above the open-charm threshold in the unit of MeV. $\Gamma^{\mathrm{theo.}}_{\mathrm{tot.}}$ denotes the summation of those in $D^{(*)}\bar{D}^{(*)},D_s^{(*)}\bar{D}_s^{(*)}$ channels which are allowed by energy conservation.}
\begin{center}\begin{tabular}{|c|c|c|c|c|c|c|c|c|}
\hline
&$\Gamma_{D\bar{D}}$&$\Gamma_{D\bar{D}^*+c.c.}$&$\Gamma_{D^*\bar{D}^*}$&$\Gamma_{D_s\bar{D}_s}$&
$\Gamma_{D_s\bar{D}_s^*+c.c.}$&$\Gamma_{D_s^*\bar{D}_s^*}$&$\Gamma^{\mathrm{theo.}}_{\mathrm{tot.}}$&$\Gamma_{\mathrm{ex.}}$~\cite{PDG}\\
\hline
$\psi(3770)$&18.4&&&&&&18.4&$27.2\pm 1.0$\\
\hline
$\psi(4040)$&2.75&21.4&56.6&6.15&&&86.9&$80\pm10$\\
\hline
$\psi(4160)$&11.7&7.98&61.4&0.0349&8.75&&89.9&$70\pm10$\\
\hline
$Y(4260)$&0.417&6.73&19.6&0.773&0.126&0.261&27.9&$55\pm19$\\
\hline
$Y(4360)$&7.78&1.23&5.53&0.0960&1.09&1.13&16.9&$96\pm7$\\
\hline
\end{tabular}\end{center}\label{tab3}
\end{table*}

In general, oscillations make $\Pi$ and relevant observables sensitive to the energy scale $E$. In view of this, we calculate the decay widths of physical states at their real mass scales, i.e., the experimental masses. The decay widths of physical states are given in Table \ref{tab3}. $\Gamma^{\mathrm{theo.}}_{\mathrm{tot.}}$ denotes the summation of those in $D^{(*)}\bar{D}^{(*)},D_s^{(*)}\bar{D}_s^{(*)}$ channels which are allowed by energy conservation. The decay widths of $Y(4260)$ and $Y(4360)$ are given under the assumption that they are the $\psi'(4S)$ and $\psi'(3D)$ respectively. The widths of $\psi(3770)$, $\psi(4040)$ and $\psi(4160)$ are dominated by the two-body open-charm decays. So for these states, $\Gamma^{\mathrm{theo.}}_{\mathrm{tot.}}$ should be comparable to the observed width of the corresponding particle. From Table \ref{tab3}, one can see that our results are consistent with the experimental data. We would like to give some explanations about the open-charm channels of $Y(4360)$ here. For $Y(4360)$ (i.e., $\psi'(3D)$ in our assignment), the open-charm decay channels with $P$ wave $D$ mesons, i.g., $D_1D$, $D^*_0D^*$, are open. The present work doesn't take into account contributions from these channels. This is a drawback of this work. We argue that the most experimentally attainable charmed mesons are $D$ and $D^*$, so they are of primary interests to us. In addition, the impacts of $D_1D$, $D^*_0D^*$ channels on the mass spectrum could be weakened by re-adjusting the model parameters.

For $\psi'(4S)$ and $\psi'(3D)$, it is interesting to find that their $\Gamma^{\mathrm{theo.}}_{\mathrm{tot.}}$ are notably smaller than those of $\psi'(3S)$ and $\psi'(2D)$.  Especially $\Gamma_{D\bar{D}}$ of $\psi'(4S)$ is less than 1 MeV. For further discussions, we estimate $\Gamma(Y(4260)\rightarrow J/\psi \pi^+\pi^-)$ using the measured data $\Gamma_{ee}Br(Y(4260)\rightarrow J/\psi \pi^+\pi^-)\approx 10$ eV~\cite{BES3}. In our assignment for $Y(4260)$, the {di-leptonic} width is $\lesssim 1$ KeV. For a larger $S-D$ mixing, the {di-leptonic} width could be even smaller. So the decay width of $Y(4260)\rightarrow J/\psi \pi^+\pi^-$ is estimated to be of $\sim 1$ MeV order. With this estimation, we obtain the ratios of the decay widths of open-charm channels to the width of $Y(4260)\rightarrow J/\psi \pi^+\pi^-$ and compare them with the upper limit given by the CLEO Collaboration in Table~\ref{tabx}. We find that all but the $D^*\bar{D}^*$ channel are lower than the upper limits. { One can see that the open-charm widths of $D\bar{D},D_s\bar{D}_s,etc$ are considerably smaller than that of $J/\psi \pi^+\pi^-$ channel, which is contrary to the naive expectation for an excited vector charmonium, and the non-observations of $Y(4260)$ in these channels may be explained. On the other hand, the ratio of $D^*\bar{D}^*$ channel is above the experimental upper limit, which means the $\psi'(4S)$ assignment for $Y(4260)$ is not consistent with experiments on every aspect. }

{In a recent paper~\cite{ChenY}, the authors analyzed the light-quark $SU(3)$ singlet and octet components of $Y(4260)$ through the $e^+e^-\rightarrow Y(4260)\rightarrow J/\psi \pi^+\pi^-$ process, and found a large octet component in $Y(4260)$. They concluded that $Y(4260)$ cannot be a conventional charmonium or a hybrid. Their work doesn't invalidate our efforts, because their results still suffer from uncertainties, and studies from different points of views, such as what we presented here, is still worthy.}

\begin{table}
\caption{The ratios $\frac{\Gamma(Y(4260)\rightarrow X)}{\Gamma(Y(4260)\rightarrow J/\psi \pi^+\pi^-)}$, where $X$ is one of the  $D^{(*)}\bar{D}^{(*)},D_s^{(*)}\bar{D}_s^{(*)}$ channels, comparing to the experimental upper bounds at $90\%$ confidence level.}
\begin{center}\begin{tabular}{|c|c|c|c|c|c|c|}
\hline
Channel $X=$&${D\bar{D}}$&${D\bar{D}^*+c.c.}$&${D^*\bar{D}^*}$&${D_s\bar{D}_s}$&
${D_s\bar{D}_s^*+c.c.}$&${D_s^*\bar{D}_s^*}$\\
\hline
Our~estimations&0.417&6.73&19.6&0.773&0.126&0.261\\
\hline
Upper~bounds~\cite{CLEO}&$<4.0$&$<45$&$<11$&$<1.3$&$<0.8$&$<9.5$\\
\hline
\end{tabular}\end{center}\label{tabx}
\end{table}

For $Y(4360)$, $\Gamma_{ee}Br(Y(4360)\rightarrow \psi(2S) \pi^+\pi^-)\approx 10$ eV~\cite{Belle2} implies $\Gamma(Y(4360)\rightarrow \psi(2S) \pi^+\pi^-)$ is of $\sim 1$ MeV order. The decay widths of $D^{(*)}\bar{D}^{(*)},D_s^{(*)}\bar{D}_s^{(*)}$ channels for $Y(4360)$ are from $\sim 0.1$ MeV to $\lesssim 10$ MeV, {which implies small branching ratios. So the tension of non-observations of $Y(4360)$ in these channels and its charmonium assignment is softened.} Using the experimental data of total widths, we show some of the branching ratios of open-charm decays {which could be tested by further experimental data}:
\begin{equation}
Br(Y(4260)\rightarrow D\bar{D})=0.758\%~~~~Br(Y(4260)\rightarrow D^*\bar{D}+c.c.)=12.2\%~~~~Br(Y(4260)\rightarrow D^*\bar{D}^*)=35.6\%
\end{equation}
\begin{equation}
Br(Y(4360)\rightarrow D\bar{D})=8.10\%~~~~Br(Y(4360)\rightarrow D^*\bar{D}+c.c.)=1.28\%~~~~Br(Y(4360)\rightarrow D^*\bar{D}^*)=5.76\%
\end{equation}

\begin{figure}[hbt]
\centering
\includegraphics[width = 0.6\textwidth]{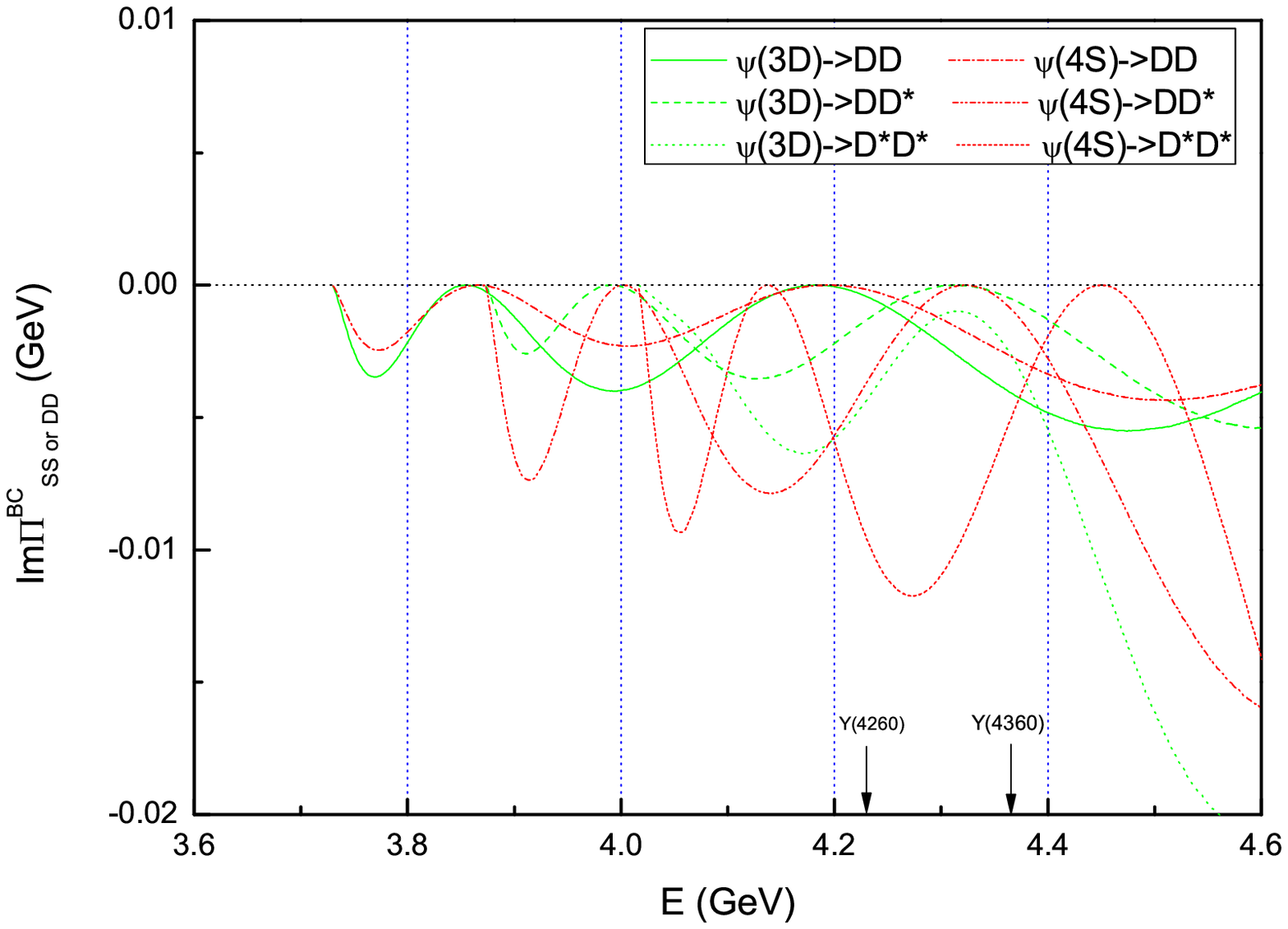}
\caption{$\mathrm{Im} \Pi^{BC}_{SS}$ and $\mathrm{Im} \Pi^{BC}_{DD}$ of $D\bar{D}$, $D\bar{D}^{*}+c.c.$ and $D^{*}\bar{D}^{*}$ channels.} \label{fig4}
\end{figure}

The reason of the smallness of decay widths of $\psi'(4S)$ and $\psi'(3D)$ are the oscillation behavior of decay amplitudes and the mixing between $S$ and $D$ wave components. The node structures of initial states make the decay amplitude of each channel oscillate, which can be reflected by $\mathrm{Im} \Pi^{BC}_{SS \mathrm{or} DD}$ in each channel. Fig. \ref{fig4} shows that the decay amplitudes (actually, the square of decay amplitudes) approach zero at particular energies. $\psi(3D)\rightarrow D^*\bar{D}^*$ dose not reach zero but is close to zero at some energy scales. From Fig. \ref{fig4}, we can see that in the range $E\simeq[4.2,4.4]$ GeV, where $Y(4260)$ and $Y(4360)$ lie in, the decay amplitude of each channel, except $\psi(4S)\rightarrow D^*\bar{D}^*$, has one trough, and this exception has one crest. So, except $\psi(4S)\rightarrow D^*\bar{D}^*$ process, $\mathrm{Im} \Pi^{BC}_{SS \mathrm{or} DD}$ of all channels at the mass scales of $Y(4260)$ and $Y(4360)$ have good chance to be small. One may wonder that the locations of the zeros of these decay amplitudes may vary as model parameters vary. But as long as we require a reasonable spectrum to be able to accommodate $Y(4260)$ and $Y(4360)$, the model parameters can not vary too far, and the locations of zeros are not sensitive to small variations in parameters.

The $S-D$ mixing then affect the final results in the following ways. For the case when the $S$ wave amplitude and the $D$ wave amplitude are not comparable to each other, for example in $D^*\bar{D}^*$ channel the $S$ wave amplitude is larger than the $D$ wave amplitude, the decay width of the mixed state dominated by $S$ wave is depressed by mixing with the smaller $D$ wave amplitude and vice versa. For the case when the $S$ wave amplitude is comparable to the $D$ wave amplitude, the decay widths of mixed states either enhanced by the constructive interference or depressed by the destructive interference. $Y(4260)\rightarrow D\bar{D}$ is an example of the constructive interference case and $Y(4260)\rightarrow D\bar{D}^*+c.c.$ is an example of the destructive interference case. In Ref.~\cite{Lla}, the authors noticed that the $S$ wave amplitude and the $D$ wave amplitude of $Y(4260)\rightarrow D\bar{D}$ decay are of opposite signs, but since the mixing angle is negative, the two amplitudes actually interfere constructively.

\begin{figure}[hbt]
\centering
\includegraphics[width = 0.6\textwidth]{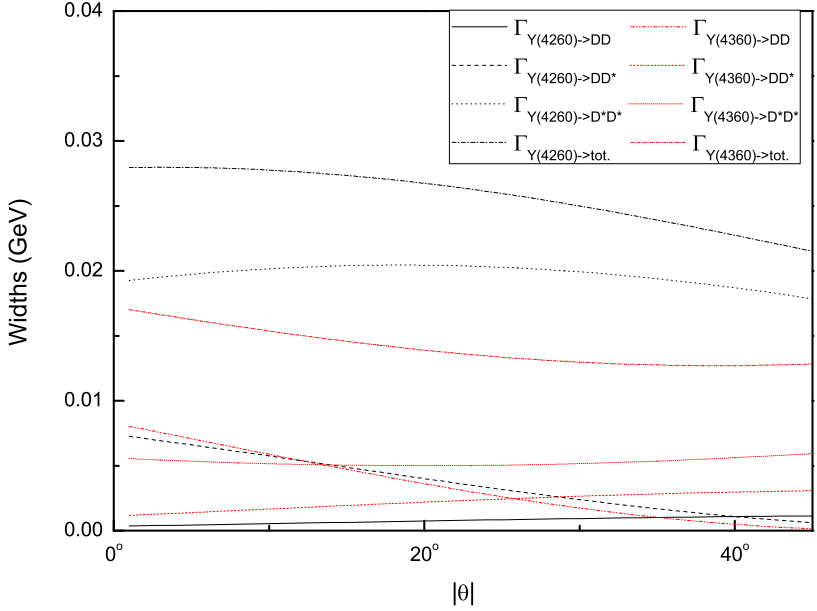}
\caption{Decay widths of $Y(4260)$ and $Y(4360)$ as functions of the mixing angle. The widths of $D_s^{(*)}\bar{D}_s^{(*)}$ are not plotted for they are generally small. The subscript "tot." in the plot means summation of the $D^{(*)}\bar{D}^{(*)},D_s^{(*)}\bar{D}_s^{(*)}$ channels.} \label{fig5}
\end{figure}

Finally, we indicate that from the {di-leptonic} decay widths of vector charmonia, one may expect larger mixing angles than the coupled-channel-induced mixing angles presented in this work. To show this we calculate the $\Gamma_{ee}$ for each physical states, and the results are shown in Table \ref{tab4}. In these calculations, the QCD correction factor $1-\frac{16}{3}\frac{\alpha_s}{\pi}$ with $\alpha_s=0.3$ is included. On the other hand, by fitting the experimental data of $\Gamma_{ee}$ under the assumption of Eqs. (\ref{SDmixing1}) and (\ref{SDmixing2}), we obtain $\theta_{2S-1D}=-11.5^{\circ}$ and $\theta_{3S-2D}=-30.7^{\circ}$. This may imply that there are some unrevealed physical sources inducing $S-D$ mixing. (The contributions in $S-D$ mixing from tensor force in potential model are smaller than coupled-channel effects~\cite{Cor2,Zou}.) { If a larger mixing angle appears in the $4S-3D$ sector, the mass splitting of $\psi'(4S)$ and $\psi'(3D)$ should become larger. In addition, the widths $\Gamma_{\mathrm{tot.}}$ for both the resonances decrease as the magnitude of the mixing angle $|\theta|$ becomes larger, as shown in Fig. \ref{fig5}. Particularly, the decay widths of $Y(4360)\rightarrow D\bar{D}$ and $Y(4260)\rightarrow D\bar{D}^*+c.c.$ decrease as $|\theta|$ increases as expected, because the $S$ wave and $D$ wave amplitudes interfere destructively in these two channels. So the measurements on the di-leptonic decays of $Y(4260)$ and $Y(4360)$ are desired.}
\begin{table} \caption{{Di-leptonic} decay widths, i.e., $\Gamma_{ee}$ in the unit of keV}
\begin{center}\begin{tabular}{|c|c|c|}
\hline
&$\Gamma_{ee}^{\mathrm{theo.}}$&$\Gamma_{ee}^{\mathrm{ex.}}$~\cite{PDG}\\
\hline
$J/\psi$&5.98&$5.55\pm0.14\pm0.02$\\
\hline
$\psi'(2S)$&2.47&$2.33\pm0.04$\\
\hline
$\psi(3770)$&0.0867&$0.262\pm0.018$\\
\hline
$\psi(4040)$&1.30&$0.86\pm0.07$\\
\hline
$\psi(4160)$&0.167&$0.48\pm0.22$\\
\hline
$Y(4260)$&0.969&\\
\hline
$Y(4360)$&0.0447&\\
\hline
\end{tabular}\end{center}\label{tab4}
\end{table}

\section{Conclusions}
{ In this work, we calculated the mass spectrum and open-charm $D^{(*)}\bar{D}^{(*)}$/$D_s^{(*)}\bar{D}_s^{(*)}$ decay widths of $J^{PC}=1^{--}$ charmonium states with the coupled-channel effects taken into account.} The mass shifts are found to be from tens MeV up to 100 MeV. We focused on the mixing between $(n+1)S$ and $nD$ states induced by the coupled-channel effects. The mixing angles are extracted. We find that the mixing in $3S-2D$ sector is larger than those in $2S-1D$ sector and in $4S-3D$ sector. { The di-leptonic decay widths are also calculated.} Most of the widths and the masses are consistent with the corresponding experimental data for states from $J/\psi$ to $\psi(4160)$. The calculations are performed using the instantaneous BS equation with the Cornell potential and the $^3P_0$ model which has been reexpressed in the form suitable for the Salpeter wave functions.

Based on these calculations, we discussed the possibility of assigning the resonant state $Y(4260)$ as the mixture of $4S-3D$ with lower mass and $Y(4360)$ as the mixture of $4S-3D$ with higher mass. { The masses of $\psi'(4S)$ and $\psi'(3D)$ are found to be $4285$ MeV and $4319$ MeV, which are close to the masses of $Y(4260)$ and $Y(4360)$ respectively, yet still deviate from them by $\sim50$ MeV. The open-charm decay widths of $\psi'(4S)$ and $\psi'(3D)$ in $D^{(*)}\bar{D}^{(*)}$ and $D_s^{(*)}\bar{D}_s^{(*)}$ channels are smaller than the naive expectations for the excited charmonia due to the oscillations of the decay amplitudes and the $S-D$ mixing effects, except for the $Y(4260)\rightarrow D^*\bar{D}^*$ channel. So the tension between the observed properties of $Y(4260)/Y(4360)$ and their conventional charmonia interpretations is softened. But the present assignments cannot be consistent with experiments on every aspect. The branching ratios of open-charm $D^{(*)}\bar{D}^{(*)}$ and $D_s^{(*)}\bar{D}_s^{(*)}$ decays and the di-leptonic decay widths are given under the present assignments, which can be tested by further experimental data.}

 \appendix
  \renewcommand{\appendixname}{Appendix}

\section{Expressions of $\mathrm{Im} \Pi_{mn}$ for Each Channel}
The decay channels we considering here are $D^{(*)}\bar{D}^{(*)}$, $D_s^{(*)}\bar{D}_s^{(*)}$, so we need to give the decay amplitudes for $^3S_1\rightarrow {^1S_0} {^1S_0}$, $^3D_1\rightarrow {^1S_0} {^1S_0}$, $^3S_1\rightarrow {^3S_1} {^1S_0}$, $^3D_1\rightarrow {^3S_1} {^1S_0}$, $^3S_1\rightarrow {^3S_1} {^3S_1}$ and $^3D_1\rightarrow {^3S_1} {^3S_1}$ processes. Now we denote the imaginary part of $\Pi_{mn}$ for each channel as $\mathrm{Im} \Pi_{mn}^\tau$, where $\tau$ represents a particular final state, i.e., $\tau={^1S_0} {^1S_0}, {^3S_1} {^1S_0}$ or ${^3S_1} {^3S_1}$. Recalling that we only consider the mixing between $nD$ and $(n+1)S$ states, so we can set $m,n$ take the value $S$ or $D$ without any confusion.
For $\tau={^1S_0} {^1S_0}$ or ${^3S_1} {^1S_0}$, $\mathrm{Im} \Pi_{mn}^\tau$ can be written as
\begin{equation}
\mathrm{Im} \Pi_{mn}^\tau (E)=- h_{m}^\tau(|\bm{P}_B|) h_{n}^\tau(|\bm{P}_B|)\times\frac{\pi|\bm{P}_B|E_BE_C}{E} \theta(E-E_{BC}^{th}),
\end{equation}
where
\begin{equation}
h_{S}^{{^1S_0} {^1S_0}}=\frac{|\bm{P}_B|\gamma}{\sqrt{3}\pi}\int \frac{d^3\bm{q}}{(2\pi)^3}R^A(|\bm{q}|) R^B(|\bm{q}-\alpha_1^B \bm{P}_B|) R^C(|\bm{q}+\alpha_2^C\bm{P}_C|)  \left(\frac{\bm{q}\cdot\bm{P}_B}{|\bm{P}_B|^2}-1\right)\frac{m_q}{\omega_{q_P}},
\end{equation}
\begin{equation}
h_{S}^{{^3S_1} {^1S_0}}=\frac{\sqrt{2}|\bm{P}_B|\gamma}{\sqrt{3}\pi}\int \frac{d^3\bm{q}}{(2\pi)^3}R^A(|\bm{q}|) R^B(|\bm{q}-\alpha_1^B \bm{P}_B|) R^C(|\bm{q}+\alpha_2^C\bm{P}_C|)  \left(\frac{\bm{q}\cdot\bm{P}_B}{|\bm{P}_B|^2}-1\right)\frac{m_q}{\omega_{q_P}},
\end{equation}
\begin{equation}
h_{D}^{{^1S_0} {^1S_0}}=\frac{-|\bm{P}_B|\gamma}{\sqrt{6}\pi}\int \frac{d^3\bm{q}}{(2\pi)^3}R^A(|\bm{q}|) R^B(|\bm{q}-\alpha_1^B \bm{P}_B|) R^C(|\bm{q}+\alpha_2^C\bm{P}_C|)  \left(1+2\frac{\bm{q}\cdot\bm{P}_B}{|\bm{P}_B|^2}-3\frac{(\bm{q}\cdot\bm{P}_B)^2}{|\bm{q}|^2|\bm{P}_B|^2}\right)\frac{m_q}{\omega_{q_P}},
\end{equation}
\begin{equation}
h_{D}^{{^3S_1} {^1S_0}}=\frac{|\bm{P}_B|\gamma}{2\sqrt{3}\pi}\int \frac{d^3\bm{q}}{(2\pi)^3}R^A(|\bm{q}|) R^B(|\bm{q}-\alpha_1^B \bm{P}_B|) R^C(|\bm{q}+\alpha_2^C\bm{P}_C|)  \left(1+2\frac{\bm{q}\cdot\bm{P}_B}{|\bm{P}_B|^2}-3\frac{(\bm{q}\cdot\bm{P}_B)^2}{|\bm{q}|^2|\bm{P}_B|^2}\right)\frac{m_q}{\omega_{q_P}}.
\end{equation}
$R^{A/B/C}$ is the radial wave function of corresponding particle. $\alpha^{B/C}_{1,2}$ is a partition parameter in defining relative momentum of quark and anti-quark of the corresponding meson. We define $\alpha_1=\frac{m_1}{m_1+m_2}$ and $\alpha_2=\frac{m_2}{m_1+m_2}$, where $m_1$ is the quark's constituent mass and $m_2$ is the anti-quark's constituent mass.

For $\tau={^3S_1} {^3S_1}$, $\mathrm{Im} \Pi_{mn}^\tau$ can be written as
\begin{equation}
\mathrm{Im} \Pi_{mn}^\tau (E)=- h_{mn}^\tau(|\bm{P}_B|) \times\frac{\pi|\bm{P}_B|E_BE_C}{E} \theta(E-E_{BC}^{th}),
\end{equation}
where
\begin{equation}
h_{SS}^{{^3S_1} {^3S_1}}=\frac{7|\bm{P}_B|^2\gamma^2}{3\pi^2}\left\{\int \frac{d^3\bm{q}}{(2\pi)^3}R^A(|\bm{q}|) R^B(|\bm{q}-\alpha_1^B \bm{P}_B|) R^C(|\bm{q}+\alpha_2^C\bm{P}_C|) \left(\frac{\bm{q}\cdot\bm{P}_B}{|\bm{P}_B|^2}-1\right)\frac{m_q}{\omega_{q_P}}\right\}^2,
\end{equation}
\begin{equation}
h_{DD}^{{^3S_1} {^3S_1}}=\frac{2|\bm{P}_B|^2\gamma^2}{3\pi^2}\frac{1}{16}\left\{8U_{D1}^2+3U_{D2}^2+U_{D3}^2+4U_{D1}U_{D2}+4U_{D1}U_{D3}+2U_{D2}U_{D3}\right\}^2,
\end{equation}
\begin{eqnarray}
h_{SD}^{{^3S_1} {^3S_1}}=h_{DS}^{{^3S_1} {^3S_1}}&=&\frac{2|\bm{P}_B|^2\gamma^2}{3\pi^2}\frac{1}{16}\{8U_{D1}U_{S1}+3U_{D2}U_{S2}+U_{D3}U_{S3}+2U_{D1}U_{S2}+2U_{D2}U_{S1}\notag\\
&&+
2U_{D1}U_{S3}+2U_{D3}U_{S1}+U_{D2}U_{S3}+U_{D3}U_{S2}\}^2,
\end{eqnarray}
and
\begin{equation}
U_{D1}=\int \frac{d^3\bm{q}}{(2\pi)^3}R^A(|\bm{q}|) R^B(|\bm{q}-\alpha_1^B \bm{P}_B|) R^C(|\bm{q}+\alpha_2^C\bm{P}_C|) \left(1-4\frac{\bm{q}\cdot\bm{P}_B}{|\bm{P}_B|^2}-3\frac{(\bm{q}\cdot\bm{P}_B)^2}{|\bm{q}|^2|\bm{P}_B|^2}
+6\frac{(\bm{q}\cdot\bm{P}_B)^3}{|\bm{q}|^2|\bm{P}_B|^4}\right)\frac{m_q}{\omega_{q_P}},
\end{equation}
\begin{equation}
U_{D2}=\int \frac{d^3\bm{q}}{(2\pi)^3}R^A(|\bm{q}|) R^B(|\bm{q}-\alpha_1^B \bm{P}_B|) R^C(|\bm{q}+\alpha_2^C\bm{P}_C|)  \left(2-2\frac{\bm{q}\cdot\bm{P}_B}{|\bm{P}_B|^2}-6\frac{(\bm{q}\cdot\bm{P}_B)^2}{|\bm{q}|^2|\bm{P}_B|^2}
+6\frac{(\bm{q}\cdot\bm{P}_B)^3}{|\bm{q}|^2|\bm{P}_B|^4}\right)\frac{m_q}{\omega_{q_P}},
\end{equation}
\begin{equation}
U_{D3}=\int \frac{d^3\bm{q}}{(2\pi)^3}R^A(|\bm{q}|) R^B(|\bm{q}-\alpha_1^B \bm{P}_B|) R^C(|\bm{q}+\alpha_2^C\bm{P}_C|) 6\left(-1+3\frac{\bm{q}\cdot\bm{P}_B}{|\bm{P}_B|^2}+3\frac{(\bm{q}\cdot\bm{P}_B)^2}{|\bm{q}|^2|\bm{P}_B|^2}
-5\frac{(\bm{q}\cdot\bm{P}_B)^3}{|\bm{q}|^2|\bm{P}_B|^4}\right)\frac{m_q}{\omega_{q_P}},
\end{equation}
\begin{eqnarray}
U_{S1}&=&\int \frac{d^3\bm{q}}{(2\pi)^3}R^A(|\bm{q}|) R^B(|\bm{q}-\alpha_1^B \bm{P}_B|) R^C(|\bm{q}+\alpha_2^C\bm{P}_C|) \left(\frac{\bm{q}\cdot\bm{P}_B}{|\bm{P}_B|^2}-1\right)\frac{m_q}{\omega_{q_P}}\frac{4}{\sqrt{2}},\\
U_{S2}&=&-U_{S1},\\
U_{S3}&=&0.
\end{eqnarray}
\section*{Acknowledgments}
{We acknowledge Dr. Yun-Hua Chen for helpful discussions. }
The work of H.-F. Fu is supported by National Science
Foundation of China (NSFC) under Grant No. 11747308 and No. 11847310.

\end{document}